\documentclass[acmsmall,screen,review]{acmart}
\renewcommand\footnotetextcopyrightpermission[1]{} 
\AtBeginDocument{%
  }
    
\usepackage{algorithm}
\usepackage{algorithmic}
\usepackage{subcaption}
\usepackage{utfsym}
\usepackage{tabularx} 
\usepackage{array}

\usepackage{multirow} 
\usepackage{booktabs} 
\usepackage{makecell} 
\usepackage{array} 
\usepackage{multirow} 

\usepackage{ragged2e} 

\usepackage{caption}
\usepackage{array} 
\usepackage{cellspace} 
\usepackage{array} 
\usepackage{cellspace} 
\usepackage{colortbl} 
\definecolor{lightgray}{gray}{0.8} 

\usepackage[utf8]{inputenc}
\usepackage{listings}
\usepackage{xcolor}

\lstdefinelanguage{Solidity}{
    keywords={contract, function, public, payable, mapping, address, uint256, msg, sender}, 
    keywordstyle=\color{cyan!80!black}\bfseries, 
    morestring=[b]", 
    stringstyle=\color{orange!80!black}, 
    morecomment=[l]{//}, 
    morecomment=[s]{/*}{*/}, 
    commentstyle=\color{gray!60}\itshape, 
    sensitive=true 
}

\begin{document}

\title{An Empirical Analysis of EOS Blockchain: Architecture, Contract, and Security}

\author{Haiyang Liu}
\authornote{Both authors contributed equally to this research.}
\affiliation{%
  \institution{Hainan University}
  \city{Haikou}
  \country{China}
}

\author{Yingjie Mao}
\authornotemark[1]
\affiliation{%
  \institution{Hainan University}
  \city{Haikou}
  \country{China}
}
\email{yingjiemao@hainanu.edu.cn}

\author{Xiaoqi Li}
\affiliation{%
  \institution{Hainan University}
  \city{Haikou}
  \country{China}}
\email{csxqli@ieee.org}

\renewcommand{\shortauthors}{Liu and Li}

\begin{abstract}
With the rapid development of blockchain technology, various blockchain systems are exhibiting vitality and potential. As a representative of Blockchain 3.0, the EOS blockchain has been regarded as a strong competitor to Ethereum. Nevertheless, compared with Bitcoin and Ethereum, academic research and in-depth analyses of EOS remain scarce. To address this gap, this study conducts a comprehensive investigation of the EOS blockchain from five key dimensions: system architecture, decentralization, performance, smart contracts, and behavioral security. The architectural analysis focuses on six core components of the EOS system, detailing their functionalities and operational workflows. The decentralization and performance evaluations, based on data from the XBlock data-sharing platform, reveal several critical issues: low account activity, limited participation in the supernode election process, minimal variation in the set of block producers, and a substantial gap between actual throughput and the claimed million-level performance. Five types of contract vulnerabilities are identified in the smart contract dimension, and four mainstream vulnerability detection platforms are introduced and comparatively analyzed. In terms of behavioral security, four real-world attacks targeting the structural characteristics of EOS are summarized. This study contributes to the ongoing development of the EOS blockchain and provides valuable insights for enhancing the security and regulatory mechanisms of blockchain ecosystems.
\end{abstract}


\maketitle
\fancyfoot{}
\pagestyle{plain} 

\section{Introduction}
With the rapid growth and widespread application of the Internet, we are entering an era of information explosion. While users benefit from the convenience of information access, they are increasingly exposed to the risk of personal data leakage. A major contributing factor to this issue is the centralized nature of most application services. In centralized systems, data security is tightly coupled with the security of platform servers. In response to these concerns, decentralized technologies such as blockchain have emerged as promising solutions for enhancing data security. Blockchain is a type of distributed database system that uses consensus algorithms to generate data, cryptographic techniques, and peer-to-peer (P2P) networks to transmit data, and a blockchain data structure to store information. The evolution of blockchain technology is generally categorized into three phases: the first phase is characterized by decentralized ledgers (e.g., Bitcoin), the second by smart contracts (e.g., Ethereum), and the third by real-world industry applications (e.g., EOS). \par
Despite the advantages blockchain offers over centralized systems, it is not inherently immune to security threats. Analyzing blockchain data remains essential for understanding and mitigating potential vulnerabilities. As blockchain-based cryptocurrencies and decentralized applications continue to grow rapidly, the vast and heterogeneous data generated by different blockchain systems presents both significant commercial value and new challenges for data analysis. Among these systems, Bitcoin, Ethereum, and EOS are widely recognized as representative platforms. \par
EOS, short for Enterprise Operating System, is considered a leading contender in the era of Blockchain 3.0. Developed by Block. One, EOSIO serves as the foundational software framework of EOS, functioning as an operating system for building blockchain infrastructure. Unlike Bitcoin and Ethereum, EOSIO introduces distinct architectural innovations in resource management, consensus mechanisms, and scalability. While there is only one official EOS mainnet, anyone can build a customized blockchain using the open-source EOSIO software. This extensibility contributes to the system's diversity and scale, with over 233 million blocks generated as of February 25, 2022. Despite its popularity, analytical research focusing on EOSIO data remains limited.\par
This paper aims to address this gap by conducting a comprehensive analysis of the EOS blockchain from three perspectives: system architecture, smart contracts, and behavioral security, leveraging relevant tools and real-world data. Although the blockchain ecosystem is rapidly evolving and EOS has emerged as a prominent Blockchain 3.0 platform, analytical studies of EOS data are still relatively sparse compared to those of Bitcoin and Ethereum. Through in-depth analysis of large-scale blockchain data, researchers can uncover valuable insights that not only hold commercial potential but also advance the development of blockchain technologies. Specifically, research on smart contracts and behavioral security can play a vital role in improving the security and regulatory oversight of blockchain systems.\par
In terms of blockchain data collection and preprocessing, Weilin Zheng et al. ~\cite{1} collected and processed EOS blockchain data and released it through XBlock (http://xblock.pro), a data-sharing platform designed to promote healthy development and research in the blockchain ecosystem. The authors organized the data into seven distinct datasets and discussed the potential research directions based on these resources.\par

Regarding system architecture, Brent Xu et al. ~\cite{2} analyzed the EOS blockchain from five perspectives: accounts, transactions and contracts, state management and blocks, execution environment, and consensus algorithms.\par

In a decentralization analysis, Adem Efe Gencer et al. ~\cite{3} conducted comparative studies on the decentralization of Bitcoin and Ethereum by evaluating network resources, analyzing node topology, and testing the robustness of these systems against attacks. Wuke Ke et al. ~\cite{4,5} proposed quantifiable methods for assessing blockchain decentralization based on entropy theory and coefficient of variation, though their evaluations also focused solely on Bitcoin and Ethereum. Jieli Liu et al. ~\cite{6} analyzed the decentralization evolution of the EOS blockchain by extracting time-series data from the DPoS network and proposed methods to detect voting collusion and abnormal mutual voting behaviors in EOSIO. Yijing Zhao et al. ~\cite{7} conducted a graph-based analysis of EOS user activity through degree distribution, clustering coefficients, and connected components to explore decentralization issues within the EOS blockchain. \par

In terms of performance evaluation, Iman Dernayka et al. ~\cite{8} conducted experimental comparisons of EOSIO and Ethereum on Microsoft Azure, evaluating response time, memory consumption, and CPU utilization. The results revealed that EOSIO exhibited longer response times. Brent Xu et al. ~\cite{2} further examined EOSIO's performance on the Whiteblock platform by measuring 13 metrics, including block size, block interval, transaction throughput (block-level and chain-level), CPU utilization, transaction types, accounts, and overall transaction volume.\par

In the domain of smart contract analysis, Suvitha M. et al. ~\cite{9} conducted a comparative study of eight major smart contract platforms—Ethereum, Hyperledger Fabric, NEM, Stellar, EOS, Waves, Corda, and Tezos—highlighting EOS’s flexibility and upgradability. Shiyi Lin et al. ~\cite{25} compared the smart contract mechanisms of Ethereum, Hyperledger Fabric, and EOS. Junzhou Xu et al. ~\cite{10} surveyed nine vulnerability detection tools applicable to Ethereum and EOSIO, summarizing their methodologies, supported vulnerability types, and limitations. Their findings indicated that existing tools for EOS smart contract auditing are limited in number, lack diversity in detectable vulnerabilities, and often cannot automate the detection process. Di Zhu et al. ~\cite{11} proposed a series of methods to measure smart contract bytecode similarity using machine learning and static analysis. Dong Wang et al. ~\cite{12} developed WANA, a cross-platform vulnerability detection tool based on WebAssembly bytecode analysis, which effectively detects three typical EOSIO vulnerabilities: fake notifications on transfers, forged EOS transfers, and predictable randomness due to block information dependency. Ningyu He et al. ~\cite{13} introduced EOSafe, a static analysis framework for EOSIO smart contracts at the bytecode level that detects four representative vulnerabilities: forged EOS transfers, forged receipts, rollback attacks, and lack of permission verification. Yuhe Huang et al. ~\cite{14} proposed EOSFuzzer, a black-box fuzzing framework designed to uncover smart contract vulnerabilities in EOSIO. Although the types of vulnerabilities it targets are the same as those detected by WANA, EOSFuzzer employs fuzz testing, while WANA supports cross-platform analysis for both Ethereum and EOSIO. \par

In the area of behavioral security, Sangsup Lee ~\cite{15} presented and investigated four types of attacks that exploit structural characteristics specific to EOSIO. Yuheng Huang et al. ~\cite{16,22} leveraged large-scale EOSIO data to detect bot activities and fraudulent behavior automatically, ultimately identifying thousands of bot accounts and multiple real-world attack patterns.\par
As evidenced by previous studies, research on the EOS blockchain system remains limited and lacks a comprehensive analysis. This paper begins with an introduction to the development of blockchain technology and the EOS blockchain, highlighting the challenges posed by the vast amount of blockchain data and the heterogeneous nature of blockchain systems, which complicates the analysis across different blockchain platforms. The security of blockchain systems cannot be overlooked, as attacks on blockchain networks can lead to significant economic losses. The significance of this work lies in its potential to enhance the security and regulatory aspects of blockchain technology. By reviewing both domestic and international literature, this paper analyzes the current state of research and summarizes the progress and achievements in the field. Furthermore, the paper provides an overview of the fundamental concepts and key technologies of blockchain, including the definition and core principles underlying blockchain systems.\par

The primary focus of this research is to conduct a systematic analysis of the EOS blockchain, utilizing data from the XBlock platform, performance analysis tools, and smart contract vulnerability detection tools. The analysis covers several critical aspects of the EOS blockchain, including system architecture, decentralization, performance, smart contracts, and behavioral security.\par

The main contributions of this study are as follows:
\begin{itemize}
\item This paper presents a comprehensive analysis of the EOS blockchain system architecture based on six core components—account management, consensus algorithm, P2P network communication, state management, transactions and contracts, and the smart contract execution environment. It provides a detailed explanation of the working principles and characteristics of each component, addressing a gap in the research on the architectural design of the EOS blockchain.
\item Based on data from the XBlock platform, this paper reveals several decentralization issues in the EOS blockchain, such as low account activity, limited participation in block producer elections, and minimal changes in the set of block producers. In addition, performance analysis shows a significant gap between EOS's actual throughput and its theoretical goal of achieving millions of transactions per second (TPS) and explores multiple factors that impact system performance.
\item This paper summarizes five common types of vulnerabilities in EOS smart contracts (such as integer overflow and fake EOS attacks), along with corresponding defense strategies. It also compares the strengths and weaknesses of four vulnerability detection tools. In addition, the paper analyzes four potential behavioral security attacks in EOS (such as block delay attacks and CPU exhaustion attacks) and proposes directions for security improvements based on real-world attack cases.
\end{itemize}

\section{Background}

\subsection{Blockchain Technology}
Blockchain, as a novel decentralized infrastructure and distributed paradigm, is a product that meets the demands of a new industry model characterized by "equality, freedom, consensus governance, and transparency." Blockchain adopts a decentralized organizational approach, achieving system organization through the interaction of distributed nodes in a bottom-up manner ~\cite{SmartBugBert}. The technology employs a consensus-based data update mechanism, which makes it extremely resistant to tampering and forgery. Furthermore, blockchain systems implement a data-reading mechanism built on privacy protection, ensuring that data remains publicly accessible while safeguarding users' privacy. As a result, blockchain is characterized by decentralization, resistance to tampering, and traceability.
Blockchain's data processing and operation rely on its layered structure. In academic research, to explore scalability solutions for blockchain, it is commonly divided into three layers: Layer 0, Layer 1, and Layer 2. \par
\begin{figure}[htp!]
	\centering
        \includegraphics[width=0.9\linewidth]{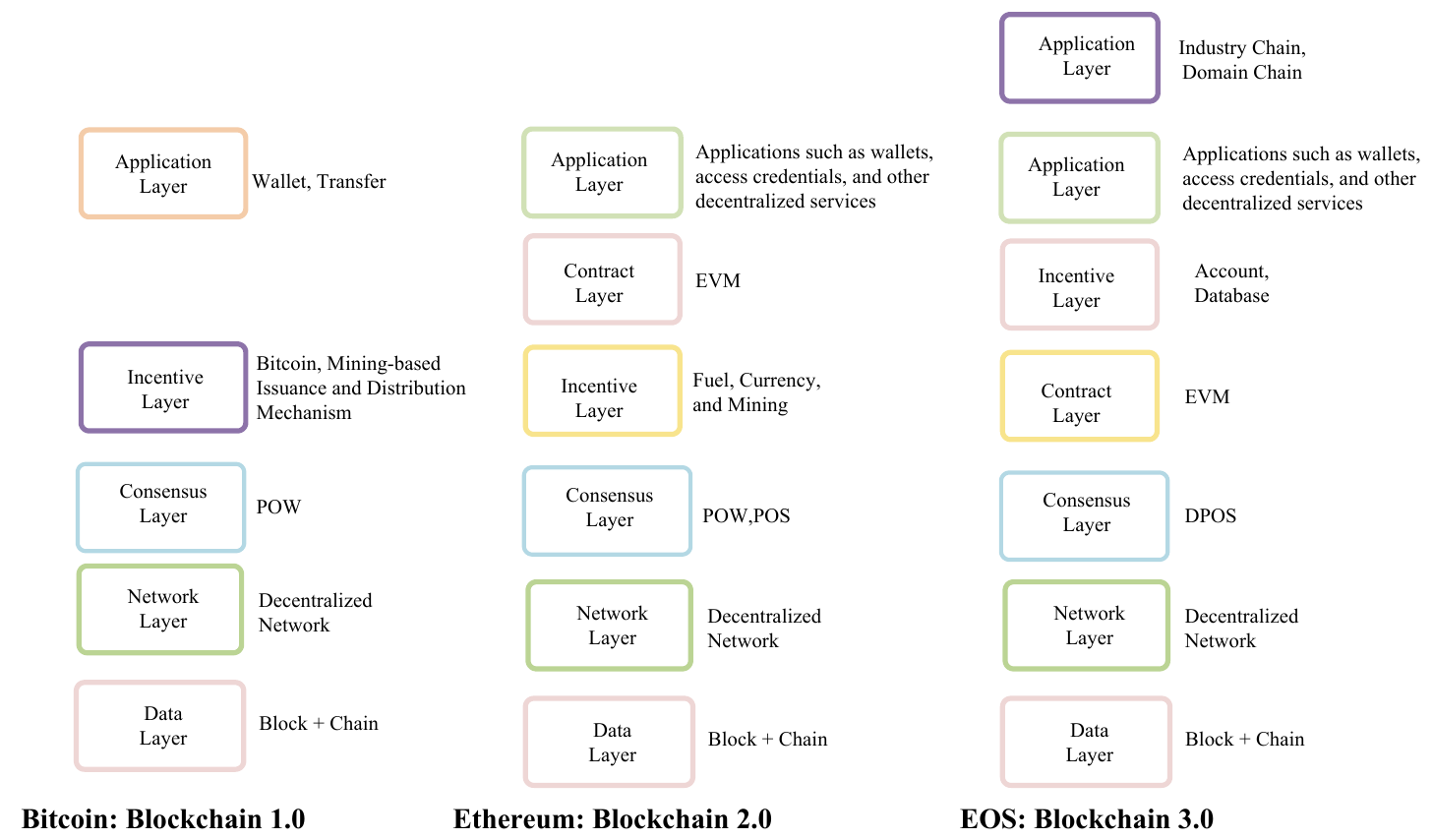}
	\caption{A comparison of the layered architectures of Bitcoin, Ethereum, and EOS.}
	\label{eos}
\end{figure}

The development of blockchain technology has undergone three distinct stages: Blockchain 1.0, Blockchain 2.0, and Blockchain 3.0. The era of Blockchain 1.0 refers to the application of virtual digital currencies, with the primary goal of achieving decentralization and enabling payment functionalities for digital currencies.~\cite{Hybrid-analysis} Blockchain 2.0 represents the phase in which blockchain technology was applied to the financial sector, marked by the advent of smart contracts ~\cite{Discovering}. The integration of smart contracts with digital currencies expanded the potential use cases of blockchain in the financial industry. Blockchain 3.0 extends the application of blockchain technology beyond the financial sector, aiming to address trust issues and ensure the security of data transmission across various industries ~\cite{statistical}. It can be said that Blockchain 1.0 represents the nascent stage of blockchain technology, Blockchain 2.0 signifies the technological implementation of blockchain in the financial and smart contract domains, while Blockchain 3.0 focuses on addressing trust and data security issues across industries. The representatives of these three stages are Bitcoin, Ethereum, and EOS, respectively. The second stage, represented by Ethereum, introduced the concept of smart contracts. The third stage, represented by EOS, employs a consensus mechanism (DPoS) that differs from the one used in Bitcoin and Ethereum. Compared to Bitcoin, Ethereum adds a contract layer between the incentive layer and the application layer, while EOS reduces the incentive layer and introduces a tool layer and an ecosystem layer ~\cite{GasTrace}. Figure \ref{eos} illustrates a comparison of the layered architectures of Bitcoin, Ethereum, and EOS.

\subsection{Consensus Mechanism}
The consensus mechanism, proposed based on the famous Byzantine Generals Problem, is designed to solve the consistency issue in distributed systems. Currently, the mainstream consensus algorithms in blockchain include PoW, PoS, and DPoS.\par
\textbf{Proof of Work (PoW).} The PoW algorithm follows the principle of "more work, more rewards." Nodes in the network solve hash functions to prove their computational effort. The first node to complete the hash calculation is rewarded with the right to record transactions. PoW relies on computational power and is mainly used in systems like Bitcoin.

\textbf{Proof of Stake (PoS).} PoS is a consensus mechanism similar to earning dividends from stock holdings. The more digital currency a participant holds and the longer they hold it, the higher their chances of earning rewards and gaining the right to validate transactions. PoS is energy-efficient compared to PoW and is often used in blockchains like Ethereum (after its transition to Ethereum 2.0).

\textbf{Delegated Proof of Stake (DPoS).} DPoS is a consensus mechanism that does not rely on computational power. Instead, it uses voting to elect block producers who act on behalf of the stakeholders to perform the consensus duties. DPoS aims to improve scalability and transaction speed by reducing the number of validators, though it may increase centralization in the network.
\subsection{Smart Contract}
Smart contracts ~\cite{SCALM} are computer programs that run on distributed ledgers, capable of automatically verifying and executing transactions without intermediaries. They emerged in the second phase of blockchain development. Current smart contract development platforms include Bitcoin scripts, Ethereum, Hyperledger Fabric, and EOS. The core of Ethereum is the Ethereum Virtual Machine (EVM), where users can develop smart contracts using mainstream programming languages or specialized languages like Solidity and Serpent. Compared to Ethereum, EOS smart contracts have several distinct differences. Firstly, the upgrade mechanism is different: EOS smart contracts are more flexible, allowing nodes to upgrade system smart contracts, and users can also update smart contracts with update permissions ~\cite{Transaction}. Additionally, each EOS smart contract is accompanied by a Ricardian contract, which describes the contract in a way that enhances readability, making it easier for parties to review and reach a consensus quickly. Another difference lies in naming conventions: Ethereum smart contracts have unique addresses, while in EOS, once a smart contract is deployed, it is bound to an account, and the contract's name becomes the account name. Finally, there is a difference in resource consumption: unlike Ethereum, which requires gas for executing smart contracts, EOS smart contracts do not require transaction fees. Instead, the bandwidth and CPU resources required are allocated based on the user's token holdings. Smart contracts deployed on the EOSIO platform consist of WebAssembly (WASM) bytecode and Application Binary Interface (ABI) ~\cite{leveraging}. The source code of the smart contract is compiled into WASM bytecode for execution in the WASM virtual machine, while the ABI describes the public interface used to interact with the smart contract. To develop a smart contract, developers must first write C/C++ files and use the `eosiocpp` compiler to compile these files into ABI, WASM, and WAST files. After debugging, these three files are uploaded to the blockchain~\cite{26}. There are two types of smart contracts on the EOSIO platform: system contracts, which are deployed by default on the platform to implement core blockchain features such as consensus and account creation, and user-defined contracts, which are deployed by users to fulfill specific business needs.

\section{Analysis of EOS Blockchain System Architecture}
\subsection{EOS Program}
Graphene is a two-dimensional carbon nanomaterial composed of carbon atoms arranged in a hexagonal honeycomb lattice via sp-hybridized orbitals. Inspired by graphene materials, EOS founder Daniel Larimer, leading the Cryptonomex team, developed the Graphene blockchain underlying technology architecture. This architecture has become an outstanding core foundational framework in the blockchain field. Graphene-based blockchains, including EOS, are characterized by fast transaction speeds, high throughput, strong stability, complete functionality, and ease of use. The underlying Graphene software architecture of EOS determines its system architecture advantages. The Graphene blockchain is not a single application program; it is composed of a series of libraries and executable programs, and it provides nodes for deploying distributed applications. One of the key technologies of the Graphene blockchain is its high modularity. The distributed communication capabilities between internal nodes are encapsulated as plugins, dynamically loaded, and called by upper-layer applications. This allows application developers to focus on the application logic without worrying about the blockchain underlying details, significantly reducing development difficulty and enhancing scalability. As a result, EOS can be seen as a collection of interacting applications within a distributed database structure.\par
\begin{figure}[htp!]
	\centering
        \includegraphics[width=0.9\linewidth]{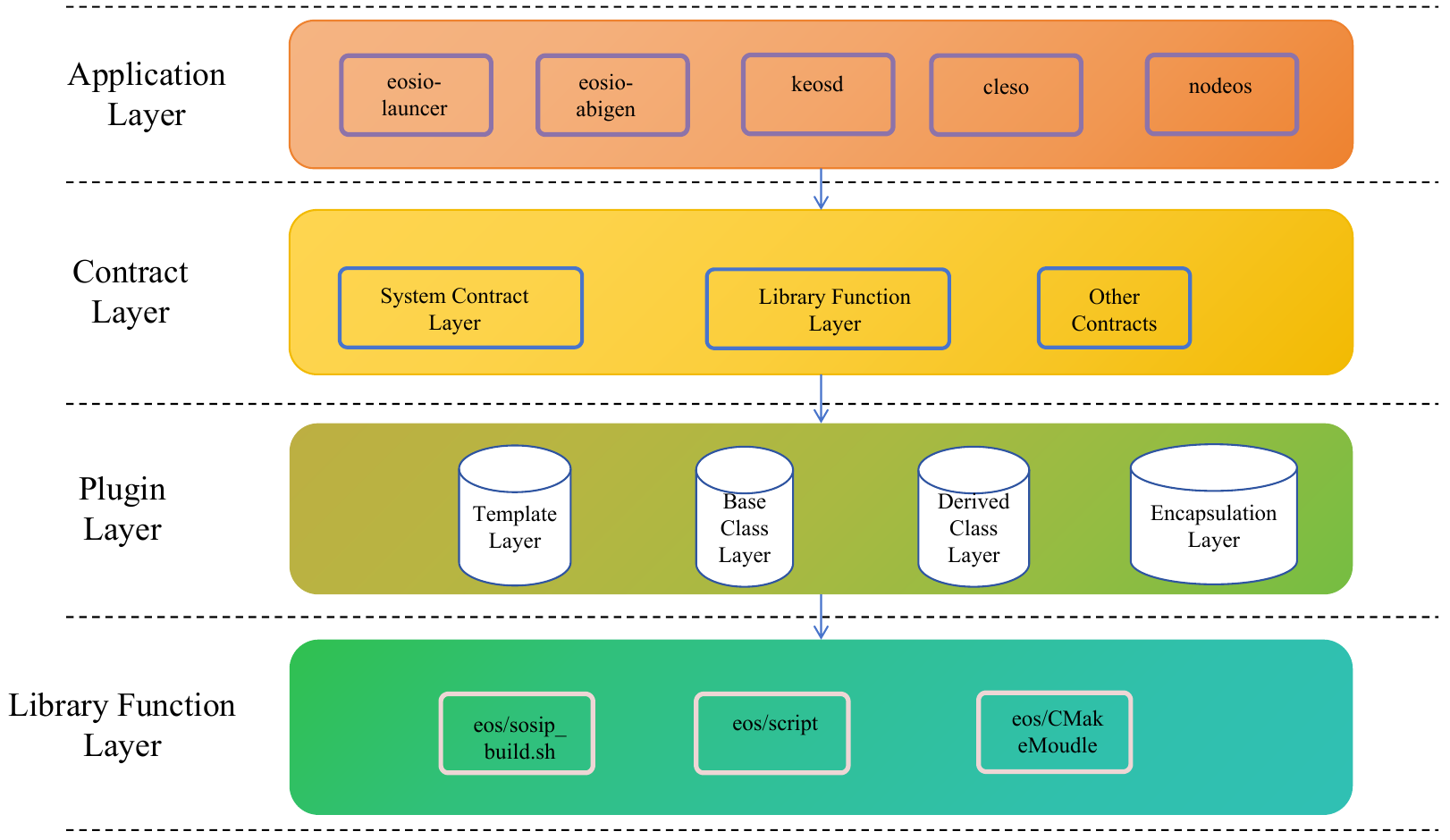}
	\caption{Code architecture of EOS.}
	\label{Code architecture of EOS}
\end{figure}

In terms of the actual EOS project, the EOS code framework can be divided into four layers, as shown below: the library function layer, plugin layer, contract layer, and application layer. The library function layer implements the key underlying technologies of the blockchain, such as transaction processing, block production, encryption functionality, file I/O operations, and network communication capabilities, providing fundamental capabilities to the application and plugin layers. The plugin layer allows different plugins to combine, and the deployment and invocation of smart contracts in the contract layer can achieve specific functionalities, offering various services. The application layer provides external service interfaces. Figure \ref{Code architecture of EOS} illustrates the code architecture of EOS. \par
The EOS system is composed of several applications, primarily including Nodeos, Keosd, and Cleos, each serving specific functions.\par
\textbf{Nodeos.} This is the blockchain node component that runs on the server side and is responsible for launching the EOS node service. It is the core process, handling account management, block generation, consensus establishment, and providing an environment for the execution of smart contracts.\par
\textbf{Keosd.} This is the wallet management program responsible for managing wallets, keys, and signing transactions.\par
\textbf{Cleos.} Cleos is a standard client program and a command-line parser with powerful extensibility. It is the command-line tool used to interact with both Nodeos and Keosd. All the data Cleos requires is retrieved through the HTTP protocol by connecting to Nodeos and Keosd, utilizing reflection mechanisms.\par
Figure \ref{The interaction process among the three programs: Cleos, Keosd, and Nodeos.} illustrates the interaction process among the three main programs.\par
\begin{figure*}[htp!]
	\centering
        \includegraphics[width=0.7\linewidth]{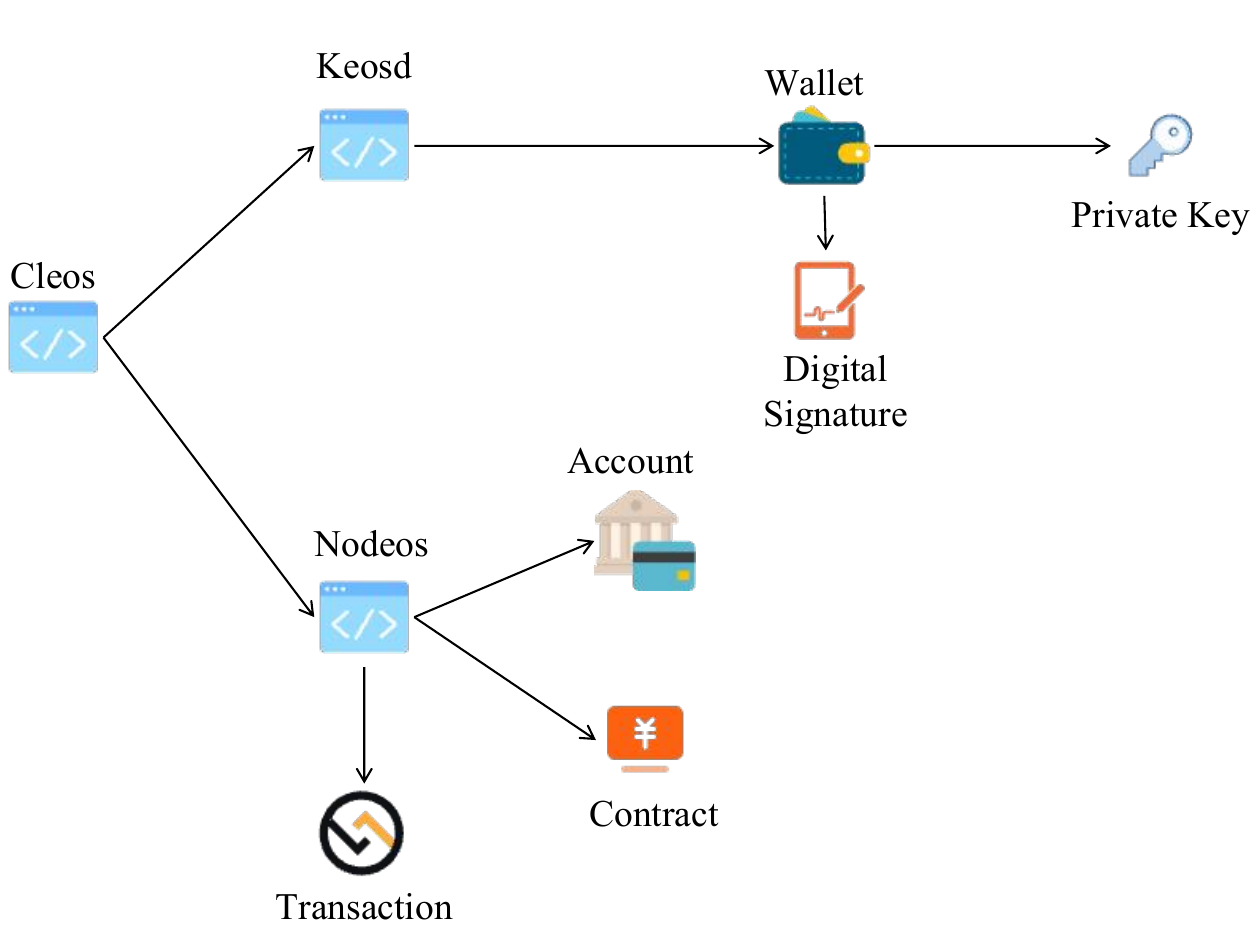}
	\caption{The interaction process among the three programs: Cleos, Keosd, and Nodeos.}
	\label{The interaction process among the three programs: Cleos, Keosd, and Nodeos.}
\end{figure*}
The EOS system is composed of six major components, which are as follows: the P2P network communication component, the state management and data block component, the consensus algorithm component, the account management component, the transaction and contract component, and the smart contract execution environment component.  This section will analyze these six components to study the EOS system architecture and its working process.
\subsection{Account Management Component}
The EOS account system is an authorization structure that features role-based access control as one of its key characteristics. Additionally, it implements an account recovery function, allowing users to create more than one account and manage them in an organized manner.\par
\subsubsection{Role-Based Permission Management}
The three essential elements of a common blockchain account are: public key, private key, and address. The address is where assets are stored, the public key represents the account name, and the private key is the complement of the public key. Typically, a pair of public and private keys represents a single account. 
In contrast, EOS accounts are identified by a 12-character human-readable identifier, which is user-defined. An EOS account can have multiple public-private key pairs, each corresponding to different permissions. This can be understood as EOS accounts being a collection of individual permissioned accounts, each represented by traditional public-private key pairs. By assigning permissions, a single account can be used by an individual or a group. All operations performed by EOS accounts are contingent on permission validation. EOS account permissions can be categorized into three types: Owner permission, Active permission, and Custom permission. Owner and Active permissions are the two native permissions, meaning they are automatically granted when the EOS account is created. For accounts with only one public-private key pair, the assets are typically stored under the public key. If the private key is lost or stolen, the account's assets become vulnerable. EOS accounts assign different weights to each entity (public-private key pair) and set thresholds for certain permissions. Only when the total weight of an individual’s keys reaches the specified threshold do they gain the corresponding permissions to perform specific operations.

\subsubsection{Account Recovery Function}
The account recovery function primarily refers to the restoration of accounts in cases of key loss or theft. In Bitcoin and Ethereum, if a private key is lost or stolen, the entire account is effectively lost. However, EOS does not follow this model; it supports the recovery of stolen keys, allowing users to regain control over their accounts and restore lost access.

\subsubsection{Account Creation Process}
Account creation and key generation occur simultaneously. Through the Cleos application, the generated keys are linked to their corresponding wallets. Cleos, Keosd, and Nodeos work together to publish accounts to the network. The wallet interacts with the network through Cleos. Keosd manages the wallet and allows users to access their accounts through Cleos.
Cleos sends an account creation request to Nodeos, which publishes the account to the network. Since private key loading is done through Cleos, the interaction between Cleos, Nodeos, and Keosd relies on the keys associated with the account for signing the transactions.
\subsection{Consensus Algorithm Component}
The purpose of the consensus mechanism is to achieve data consistency across nodes in a distributed environment. Unlike Ethereum, which uses PoW, EOS employs DPoS as its consensus mechanism. Later, EOS improved this mechanism by incorporating the Byzantine Fault Tolerance (BFT) algorithm, resulting in the BFT-DPoS consensus mechanism.
\subsubsection{DPoS and BFT-DPoS}
The DPoS consensus mechanism involves two phases: the Voting Phase, where token holders elect 21 trusted block producers (BPs) to validate transactions and produce blocks, and the Block Production Phase, where these BPs follow specific rules to produce, broadcast, and confirm blocks.\par
In the BFT-DPoS consensus mechanism, when a block producer (BP) produces a block, it is still broadcast to the entire network. However, unlike in traditional DPoS, other BPs immediately validate and confirm the new block upon receipt rather than waiting until their turn to produce a block. As a result, the transaction confirmation time is reduced from 45 seconds to 3 seconds.\par
\subsubsection{Consensus Algorithm Workflow}
In the EOS system, a block cycle consists of the production of 252 blocks. At the beginning of each block cycle, EOS conducts a vote based on the EOS tokens held by users across the network. According to the voting results, 21 supernodes are selected. These supernodes, also known as block producers (BPs), are responsible for collecting transaction information and packaging it into blocks. When a block producer generates the next block, they are required to execute and verify the transactions and contracts that clients in the network send to them. The block producers process valid transactions and contracts while filtering out invalid ones. Once the block producer has processed the transactions, it broadcasts the information to other block producers in the network. Once the majority (more than two-thirds) of the 15 block producers (out of the total 21 block producers) validate and sign off on the dataset, the transaction is considered finalized. The 21 supernodes agree on the block production order, and a new block is produced every 0.5 seconds according to this order. If a supernode fails to produce a block on time, it is skipped and given a timestamp. If this supernode fails to produce a block within 24 hours, it will be removed from the list of candidate supernodes.\par
\subsection{P2P Network Communication Component}
A blockchain system's P2P module should be capable of synchronizing block data from peer nodes, sending transactions to other nodes for validation, validating transactions sent by other nodes, broadcasting blocks generated by the node to others, and verifying blocks received from other nodes. This section will analyze the functionality of the P2P module through the net plugin.\par
\subsubsection{net\_plugin Class}
The plugin class is responsible for managing the registration, initialization, and state of plugins. It provides the functionality to initialize, start, and stop plugins within the system. The net$\_$plug in class, as a subclass of the plugin class, specifically implements the configuration, initialization, startup, shutdown, and block broadcasting functionalities for the P2P network plugin. Through the plugin\_startup function, a P2P node is started, and a listening loop is set up to respond to messages received from other nodes. Additionally, based on the seed node information in the configuration file, the node connects to other peers, sends message requests, and synchronizes blocks or other network data. This allows the net\_plugin to handle critical networking tasks, ensuring that the node can communicate efficiently with other nodes in the EOS network, propagate blocks and transactions, and maintain synchronization with the rest of the blockchain.
\subsubsection{Block Synchronization Process in EOS}
After the local node starts, it connects to the P2P nodes specified in the configuration file and retrieves the current blockchain information and configuration data. It then constructs a handshake\_message packet and sends it to other P2P nodes. When a remote node receives this message and detects that its chain is longer than the local node's chain, it sends a notice\_message to the local node to initiate block synchronization. Upon receiving this message, the local node checks the block synchronization information and constructs a sync\_request\_message, which is then sent to the remote node to request block data. After receiving the request, the remote node fetches its n-th block as requested by the local node, constructs a signed\_block message containing the block data, and places it into the message queue. This way, the requested block is sent to the local node. During this process, the remote node sends blocks recursively based on the local node’s requests, sending one block at a time. The local node receives each signed\_block message—which contains the detailed data of the block, and completes the block synchronization accordingly.

\subsection{State Management and Data Block Components}
State represents information and events and is closely related to system availability and stability. Therefore, the effective management of system states and resources is critical. This section analyzes three key aspects of the EOS blockchain system: resource management, state management, and data management.

\subsubsection{Resource Management}
The EOS system mainly utilizes three types of resources: bandwidth and log storage, computation and computational backlog (CPU), and state memory (RAM). EOS adopts a tokenized resource allocation mechanism, where resources are distributed based on the proportion of tokens held by users.\par

\textbf{Bandwidth and Log Storage.}
Block producers need to synchronize generated blocks with other producers, which consumes bandwidth. Bandwidth can be acquired in two ways. If a user holds tokens, they can stake them to a system account and receive bandwidth proportionally based on their share of the total token supply. Alternatively, users without tokens can lease bandwidth. Each transaction consumes a certain amount of bandwidth. When the bandwidth is depleted, the user can no longer send transactions.

\textbf{Computation and Computational Backlog (CPU).}
Calling and executing smart contracts consumes CPU resources, which are measured by the time taken to execute contract code. CPU can be obtained by staking tokens or purchasing it from other users. Each contract invocation consumes CPU; when exhausted, further execution of smart contracts becomes impossible.

\textbf{State Memory (RAM).}
RAM refers to runtime memory used to store account-related state information. Compared to bandwidth and CPU, RAM is a more scarce resource, with its total amount determined collectively by the block producers. RAM is also obtained via token staking; however, unlike other resources, RAM staking and unstaking incur a 0.5\% fee. RAM allocated through staking is non-transferable and cannot be leased or sold to other users. While staking and unstaking RAM are immediate operations, token staking for CPU and bandwidth comes with a 3-day lock period. CPU and bandwidth are renewable over time, whereas RAM is a fixed resource and must be repurchased once consumed.

\subsubsection{State Management}
Transactions, smart contract invocations, and executions in the EOS blockchain all lead to state changes. EOS ensures state persistence through a database management system infrastructure. This persistence is crucial for maintaining the history and integrity of transactions. Modifying the application state requires write access and must follow a predefined set of rules. Reconstructing application state from log data through logical validation involves three steps: verifying internal consistency, checking that preconditions are satisfied, and finally, applying the state change.
During this process, read-only operations can be executed in parallel, but write operations must be executed sequentially.

\subsubsection{Data Management}
Throughout the lifecycle of a transaction, the EOS system employs various data structures to store blockchain data. Four data access modes control how nodes process transactions, blocks, and messages when Nodeos is run in different configurations. Nodeos stores historical transactions, transaction records, and current states. Historical transaction data is saved in the blocks.log file on disk. Current state data is stored in either Chainbase or RocksDB. Chainbase is a private in-memory transactional database constructed by blocks. It achieves persistence using memory-mapped files. RocksDB is an open-source, persistent key-value store optimized for flash drives and high-speed disks, utilizing in-memory caching for efficient data access.
EOS also provides a persistent data structure known as the Multi-Index Table, which is a set of classes for database operations that support persistent storage.\par
There are four data read modes in EOS: Speculative Mode, Head Mode, Read-Only Mode, and Irreversible Mode. The current data access mode is specified via the --read-mode option in eosio::chain\_plugin.

\subsection{Transaction and Contract Components}
In the EOS system architecture, transactions, actions, and contracts are closely interrelated. Understanding the transaction and contract components requires a clear grasp of the relationship among these three entities. An action is the atomic execution unit on the EOS blockchain. Most EOS network operations are built upon actions, which are capable of interacting with each other. A set of actions, along with their corresponding action handlers, constitutes a smart contract. A transaction represents the execution of one or more actions and serves as the carrier for delivering and invoking them.
\subsubsection{Contract Dispatch}
In EOS, smart contracts are programs that are registered on the network and executed on the nodes. Communication between contracts is asynchronous. The process of dispatching a contract in EOS proceeds as follows: \par
First, the client, via the Cleos component, sends a request to the Nodeos service. Nodeos then generates an action request and compiles it into WebAssembly (WASM) bytecode. During this process, a series of actions is bundled together in a defined sequence to form a transaction. Once the compilation is completed, the resulting WASM bytecode activates and enforces the corresponding contract logic.
\subsubsection{Two Communication Models of Transactions}
A transaction in EOS is composed of a sequence of actions that are combined and executed in order. Based on execution timing, transactions can be categorized into immediate transactions, which are executed without delay and broadcast to the network upon initiation, and delayed transactions, where execution is scheduled for a later time. EOS supports two fundamental communication models for transactions: inline communication and deferred communication. Inline communication involves invoking nested actions within a transaction by directly requesting other operations without generating any notifications outside the scope of the current transaction. Deferred communication, in contrast, sends action notifications to peer transactions. However, from the perspective of block producers, deferred actions are not guaranteed to be executed. The system only ensures whether the transaction is successfully created and submitted, without guaranteeing its eventual execution.
\subsection{Smart Contract Execution Environment Component}
A smart contract is a collection of programmatic instructions executed by block producers upon the fulfillment of specified triggering conditions. The execution of a smart contract can be conceptualized as a process where multiple inputs collectively determine the output. In EOS, a smart contract consists of two main components: a set of actions and type definitions. The action set defines and implements the behavior and functionality of the contract, while the type definitions specify the necessary content and data structures required by the contract. Two mainstream approaches for executing smart contracts are virtual machines and containers. A virtual machine interprets and executes each contract instruction at the system level, thereby providing a virtual hardware platform. In contrast, the container-based approach offers an isolated sandbox execution environment by leveraging operating system-level resource partitioning and isolation mechanisms. EOS employs the mainstream C++ programming language for writing smart contracts. Once written, these contracts are compiled into low-level WebAssembly (WASM) bytecode for execution by block producers. The execution model is based on a virtual machine architecture, which allows compatibility with multiple virtual machine implementations. The EOS blockchain system currently supports three WASM execution engines: WABT, Binaryen, and WAVM. The principles behind these three execution engines are as follows: WAVM improves virtual machine performance by precompiling the WASM instruction set into native machine code for direct execution on local hardware; however, its just-in-time (JIT) compilation introduces latency that may cause it to miss block production deadlines. Binaryen translates input into an intermediate representation (IR) for interpretation. Although WABT only supports input in WASM binary format, it effectively reduces block production time and overall system latency.
\section{Decentralized analytics}
Decentralization is an essential characteristic of blockchain technology, and its applications and advantages are primarily reflected in this decentralization. However, measuring the degree of decentralization in a blockchain system is challenging, and there are few standardized metrics for quantifying decentralization across different blockchain networks. In this section, we will conduct an analysis of the EOS blockchain decentralization by examining account data, the voting process, and block producer behaviors based on data provided by Xblock. This will allow us to gain a deeper understanding of decentralization in the EOS blockchain.

\subsection{Account Analysis}
Unlike most blockchain systems, where account creation is free, EOS requires the account creator to purchase state storage (RAM) to store account information. To initiate transactions, these new accounts also need to be allocated CPU and bandwidth resources. These resources on the EOS blockchain must either be staked with Tokens or rented from other users. To analyze accounts, we need to examine the "account" and "token" datasets from Xblock. The account data in XBlock includes information such as the account creation time, creator, and account name. According to statistics, from June 9, 2018, to September 3, 2020, a total of 2,034,330 accounts were created by 49,072 creators.\par
\begin{figure*}[htp!]
    \centering
    \begin{subfigure}{0.5\textwidth}
        \centering
        \includegraphics[width=\textwidth]{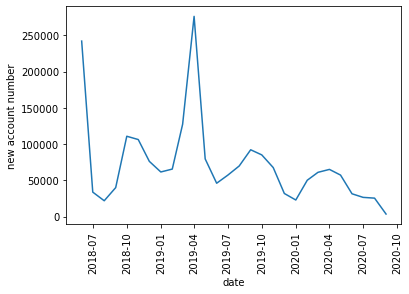}
        \caption{Account Number–Time Statistics.}
        \label{Account Number–Time Statistics}
    \end{subfigure}
    \begin{subfigure}{0.5\textwidth}
        \centering
        \includegraphics[width=\textwidth]{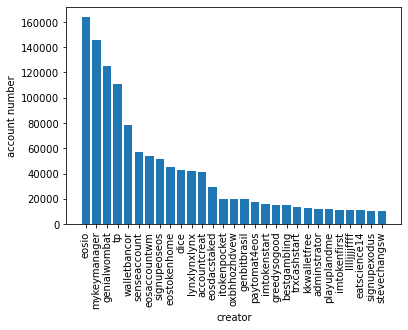}
        \caption{Distribution of Users with Account Numbers Greater than 10,000.}
        \label{Distribution of Users with Account Numbers Greater than 10,000.}
    \end{subfigure}
\end{figure*}

As shown in Figure \ref{Account Number–Time Statistics}, the number of new accounts created during each period from June 2018 to September 2020 is presented. It can be observed that the creation of new accounts on the EOS blockchain was significantly higher in June 2018 and April 2019 compared to other periods. On June 9, 2018, the EOS mainnet was officially launched, which marked a surge in account creation at the early stages of the network's launch. In 2019, the DApp industry saw rapid development, which in turn drove the growth of blockchain platforms such as Ethereum and EOS. During this period, the activity level of EOS accounts significantly increased. In 2019, EOS had over 572,000 total users, with approximately 518,000 active users. There were 493 observable DApps, of which 479 were active. \par
These 2,034,330 accounts were created by 49,072 creators, with an average of 41.46 accounts created per creator. However, because the creation of EOS blockchain accounts requires the purchase of network resources, in reality, only 2,266 creators had more accounts than this average. As shown in Figure \ref{Distribution of Users with Account Numbers Greater than 10,000.}, the account information of the 29 creators who had more than 10,000 accounts is selected. From the Figure \ref{Distribution of Users with Account Numbers Greater than 10,000.}, it can be seen that the system account "eosio" ranks first. However, even this account's number of created accounts accounts for only 8\% of the total. EOS blockchain account names can be up to 12 characters long and may use digits [1-5] or lowercase letters [a-z]. The accounts created by the system account "eosio" generally have similar names, many of which start with hexadecimal characters, indicating that these accounts are used for testing purposes.\par
\begin{figure}[htp!]
	\centering
        \includegraphics[width=0.7\linewidth]{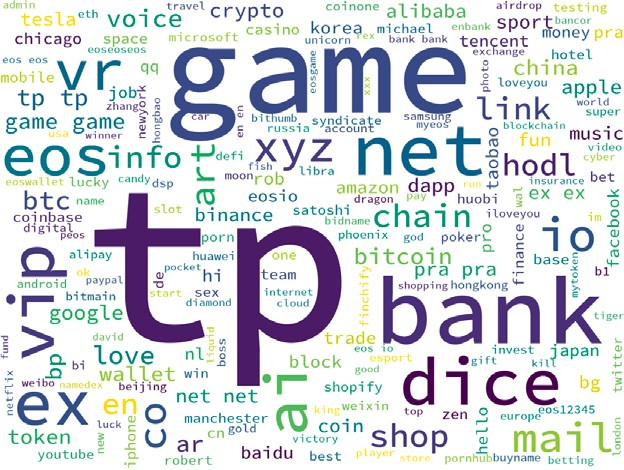}
	\caption{Account Name Word Cloud.}
	\label{Account Name Word Cloud}
\end{figure}

A word frequency analysis was conducted on all the account names to generate a word cloud, as shown in Figure \ref{Account Name Word Cloud}. The top three most frequent terms in the account names are "tp," "game," and "bank." Among these, "tp" refers to TokenPocket, a company that provides wallet services. The presence of terms like "game," "bank," and "dice" indicates that EOS applications are heavily oriented towards sectors such as banking, gaming, and gambling.
To study these accounts, it is also important to analyze their token holdings, specifically the EOS tokens that these accounts own. In EOS, a valid token contract must include at least three components: token creation, issuance, and transaction transfers. Therefore, the XBlock dataset is divided into these three sections. 
Through the analysis of the token dataset from June 2018 to September 2020, a total of 4,811 tokens were created and issued, generating 1,128,111,142 transaction transfer records. In total, 1,295,389 accounts held token assets.\par 
This indicates that 63.67\% of all accounts hold token assets, meaning one-third of the accounts do not hold tokens. This suggests that these accounts are not active in the EOS ecosystem and will not participate in the election of block producers. The next section will discuss whether these token-holding accounts are involved in the voting process.
\section{Performance Analysis}
With the development of blockchain technology, more and more blockchain system platforms have emerged. The two main indicators that determine the development of these blockchain systems are performance and security. This section will examine the performance of the EOS blockchain system from two aspects. First, it will analyze the data from XBlock, focusing on block production, transaction numbers, and resource management. Since the XBlock data in this section is only available until November 2019, and blockchain development is rapidly evolving, this section will also analyze the recent performance of the EOS blockchain using the Titan Labs (EOS) tool to examine the EOS mainnet transaction status and CPU utilization.
\subsection{The Analysis of XBlock Data}
The XBlock block dataset contains detailed data for 90 million EOS blocks, including information such as block number, creation time, producer, CPU and NET usage, as well as the number of transactions and actions. These data are divided into three parts to analyze the EOS blockchain system from the perspectives of block production, transaction and action volume, and CPU and NET usage.

\begin{figure}[htp!]
	\centering
        \includegraphics[width=0.7\linewidth]{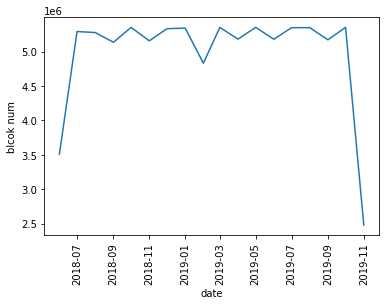}
	\caption{Block Production Over Time.}
	\label{Block Production Over Time}
\end{figure}

\begin{figure}[htp!]
	\centering
        \includegraphics[width=0.7\linewidth]{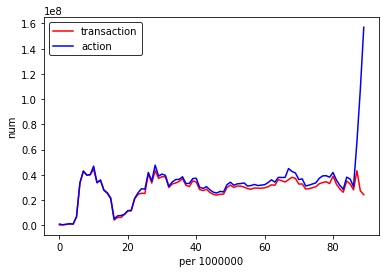}
	\caption{Statistics of Transaction and Action Counts.}
	\label{Statistics of Transaction and Action Counts}
\end{figure}

As shown in the Figure \ref{Block Production Over Time}, the monthly block production over time on the EOS blockchain is visualized. The block counts for June 2018 and November 2019 are relatively low because data for these months is incomplete. From July 2018 to October 2019, block production remained relatively stable, fluctuating between 4,831,452 and 5,351,856 blocks per month. This variance is within an acceptable range considering the differing number of days in each month and aligns with the expected block interval of approximately 0.5 seconds per block, indicating that block producers were fulfilling their responsibilities.
In terms of transactions and actions, the 90 million blocks recorded a total of 2,538,345,954 transactions and 2,960,718,844 actions. By aggregating the number of transactions and actions for every one million blocks, we obtain the results shown in Figure \ref{Statistics of Transaction and Action Counts}. As shown in Table 3, the average number of transactions per block is approximately 28.20. Based on the 0.5-second block interval, the average transaction throughput of the EOS system is about 56.40 TPS. When the block height exceeded 87.7 million, the throughput peaked at around 126 TPS. However, this is still far from EOS’s originally claimed throughput of one million TPS.

\begin{figure}[htp!]
	\centering
        \includegraphics[width=0.7\linewidth]{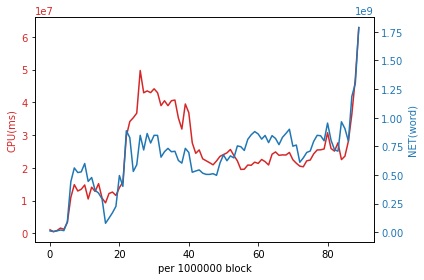}
	\caption{CPU and NET Usage Statistics.}
	\label{CPU and NET Usage Statistics}
\end{figure}

The sudden surge in transaction and action volume during this phase can be attributed to the EIDOS project, which launched a massive airdrop on EOS. This resulted in about 80\% of on-chain transactions being related to this airdrop, causing a spike in CPU resource usage on the EOS mainnet and making it difficult for normal transactions and DApp users to operate properly.
It should be noted that this analysis is based solely on the data collected from XBlock and does not represent the current real-world transaction throughput of the EOS system. A separate analysis of the current EOS transaction throughput will be conducted later.
Additionally, it is observed that the number of transactions and actions is relatively close in the figures. Since a transaction is composed of one or more actions, the total number of actions is slightly higher than the number of transactions. The fact that their counts are similar implies that most transactions consist of only a single action.
Regarding CPU and NET usage, the statistics are presented in Figure \ref{CPU and NET Usage Statistics}, where CPU usage is measured in milliseconds, and NET usage is measured in words (1 word = 8 bytes). It can be seen that the trends of CPU and NET usage closely follow the transaction volume, as transaction processing consumes these computing resources and leads to fluctuations in their usage levels.

\subsection{Performance Analysis of the Current EOS System}
\begin{figure}[htp!]
    \centering
    \begin{subfigure}{0.3\textwidth}
        \centering
        \includegraphics[width=\textwidth]{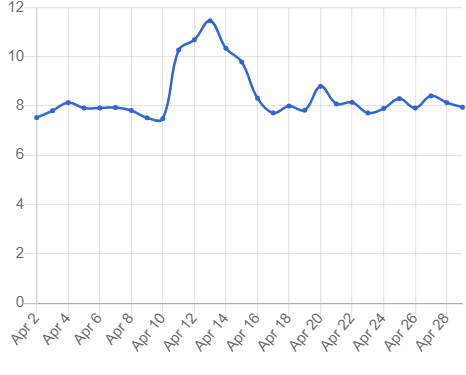}
        \caption{Average Transactions Per Second (TPS).}
        \label{Average Transactions Per Second (TPS)}
    \end{subfigure}
    \begin{subfigure}{0.3\textwidth}
        \centering
        \includegraphics[width=\textwidth]{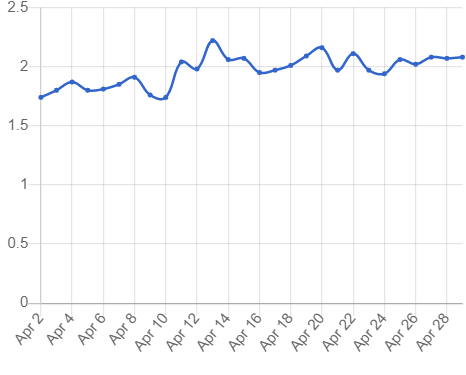}
        \caption{Average CPU Utilization (\%).}
        \label{Average CPU Utilization.}
    \end{subfigure}
        \begin{subfigure}{0.6\textwidth}
        \centering
        \includegraphics[width=\textwidth]{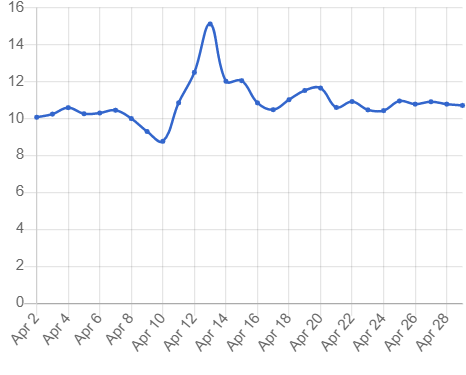}
        \caption{Average NET Usage (MB).}
        \label{average NET Usage (MB).}
    \end{subfigure}
        \begin{subfigure}{0.9\textwidth}
        \centering
        \includegraphics[width=\textwidth]{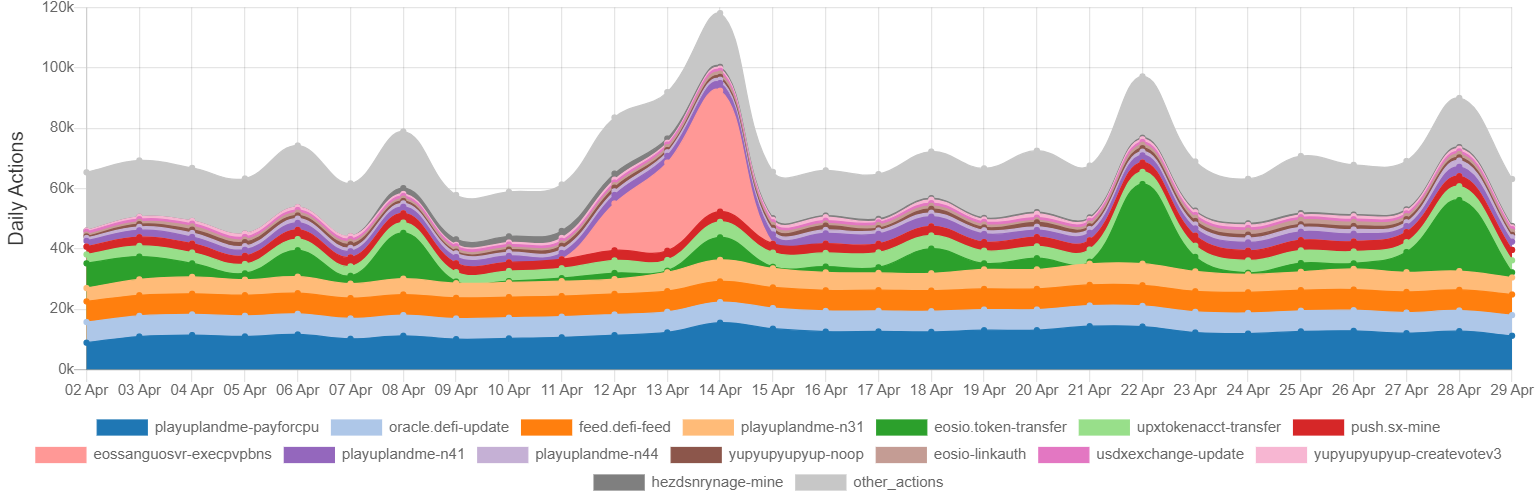}
        \caption{Average Daily Action Count.}
        \label{Average Daily Action Count}
    \end{subfigure}
     \caption{Statistical Results.}
    \label{Statistical Results}
\end{figure}

The previous section analyzed block data collected from XBlock. However, since the dataset only includes the first 90 million blocks, ending in November 2019, it does not reflect the current performance of the EOS system.
In this section, we utilize Titan Labs (EOS) to analyze the EOS mainnet’s transaction and action activity, as well as CPU and NET usage over the past 28 days. Titan Labs (EOS) is a real-time monitoring platform for the EOS mainnet, providing access to up-to-date operational metrics.
The statistical summary as of April 30, 2022, is shown in Figure \ref{Statistical Results}.

\section{Smart Contract Analysis}
Smart contract security is a critical component in the application of blockchain systems. The complexity and code size of smart contracts are increasing year by year, and the reuse of code has led to the wider distribution of contract vulnerabilities and malicious code. Once a smart contract vulnerability is exploited, it can lead to severe consequences and financial losses. As more and more blockchain platforms support the use of smart contracts to build decentralized applications, the economic damage caused by smart contract vulnerabilities to clients and users is also increasing. Therefore, analyzing smart contracts on blockchain platforms is essential. \par

Since most EOSIO smart contracts are not open source, and there are few analysis tools available for analyzing Wasm bytecode, detecting vulnerabilities in EOSIO smart contracts becomes difficult. While analysis tools for Ethereum smart contracts have developed rapidly, these tools are not suitable for EOSIO due to differences in virtual machines, bytecode structures, and vulnerability types. The challenges of analyzing EOSIO smart contracts mainly focus on the following aspects: First, the EOSIO virtual machine execution engine is more complex than Ethereum's in terms of both quantity and variety. Second, Wasm bytecode is more complex than Ethereum's bytecode, which adds difficulty to the analysis. At the same time, the types of vulnerabilities in EOSIO smart contracts are more complex than those in Ethereum, further complicating the analysis and vulnerability detection of EOSIO smart contracts.
\subsection{Common Vulnerabilities and Defense Methods of EOSIO Smart Contracts}
In this section, five known typical vulnerabilities in EOSIO smart contracts will be introduced, along with their defense mechanisms.  \par
\textbf{Integer Overflow}: Integer overflow vulnerabilities primarily occur when a contract performs arithmetic operations where the result may exceed the number of bytes the underlying computer system can store. If boundary checks are not performed, there is a possibility of integer overflow. Below is an example of code for batch transfers, where integer overflow occurs.\par
\textbf{Defense Method}: Before performing arithmetic operations, it is essential to check whether the result will exceed the allowed numerical range of the computer system.\par
\textbf{Permission Validation}: This refers to the need for permission validation within the contract, ensuring that the caller has the appropriate rights to execute the operation. Smart contracts without permission checks are vulnerable to being called by malicious accounts, which could perform harmful actions, such as modifying databases. Below is an example of a transfer process code, where the absence of permission validation leads to a mismatch between the account transferring the assets and the account calling the contract. Malicious accounts can exploit this to transfer assets that do not belong to them.\par
\textbf{Defense Method}: Developers can use require\_auth(account), require\_auth2(account, permission), and has\_auth(account) to verify the consistency of the caller and the account and ensure the necessary permissions are in place.\par
\textbf{Fake EOS Attack}: In EOSIO, a smart contract must send EOS tokens to another smart contract through the system contract `eosio.token`. During a transaction, the sender contract must invoke the `transfer` function within `eosio.token`, which adjusts the account balances of both the sender and the receiver. When the transaction occurs, `eosio.token` calls the `require\_recipient()` function, notifying both the sender and receiver. A fake EOS transfer attack refers to when a malicious account impersonates `eosio.token` and sends a `transfer` to the victim account. If the victim doesn't verify the received `transfer, ' they may mistakenly believe that the malicious account has transferred EOS tokens to them. Below is an example of code that could potentially send a forged EOS transfer. In the `switch(action)` section, if the action is calling the `transfer` function, and there is no check to verify that this transfer comes from `eosio.token`, then the attacker can invoke the vulnerable contract's `transfer` and perform the transaction without spending their own EOS tokens.
\textbf{Defense Method}: Perform a check on the action to ensure that if it is a `transfer`, it must come from `eosio.token`. Below is an improved version of the code to address this issue.\par

\textbf{Fake Transfer Notification}: During a transaction, `eosio.token` calls the `require\_recipient()` function to notify both the sender and the receiver. If the recipient account is not checked to verify whether it is the intended account, the system could mistakenly assume that the sender has transferred EOS tokens to themselves, even though the account has not received the transfer. This is known as a fake transfer notification. Below is a part of the code where a fake transfer notification attack could occur.\par

\textbf{Defense Method}: Check if the recipient of the transfer notification is indeed the correct account (i.e., the receiver). The following is an improved version of the code to mitigate this issue.\par

\textbf{Non-Random Random Numbers}: In blockchain, it is difficult to obtain a reliable source of randomness. Random number generation in smart contracts should not be controllable or predictable.\par

\textbf{Defense Method}: True random numbers cannot be generated on the EOS blockchain. It is recommended to refer to the examples provided in the official documentation when designing applications that rely on randomness.\par

\subsection{Introduction and Comparison of Smart Contract Vulnerability Detection Tools}
This section introduces four EOSIO smart contract vulnerability detection tools- EVulHunter, EOSFuzzer, WANA, and EOSafe- and compares them.\par

\textbf{EVulHunter} is a binary-level vulnerability detector for EOS smart contracts based on the Octopus project. Octopus is a security analysis framework for WebAssembly modules and blockchain smart contracts. It provides a simple way to analyze closed-source WebAssembly modules and smart contract bytecode to gain a deeper understanding of their internal behavior. Octopus supports the analysis of WASM, BTC scripts, Ethereum smart contracts (EVM bytecode), Ewasm, EOS smart contracts (WASM), and NEO smart contracts (AVM bytecode). Based on Octopus, EVulHunter can detect two smart contract vulnerabilities: fake EOS attacks and fake transfer notifications.\par

\textbf{EOSFuzzer} is a fuzz testing tool that detects EOSIO smart contract vulnerabilities in four steps. First, EOSFuzzer performs static analysis on the smart contract bytecode and its binary program interface (ABI). Based on the analysis results, it fuzzes the inputs to these interfaces according to the ABI data types and conducts attacks. The third step involves executing fuzzing on the smart contract using the Cleos tool on a testnet. Finally, vulnerabilities are detected and analyzed based on the fuzzing results. EOSFuzzer supports detecting three types of attacks: fake EOS attacks, fake transfer notifications, and block dependency. Block dependency refers to the existence of a feasible path from block information collection to EOS transfer within the smart contract, which compromises the randomness of the entire process, making the random result traceable to a deterministic outcome through block dependency.\par

\textbf{WANA} is a cross-platform tool that supports the detection of vulnerabilities in both EOSIO and Ethereum smart contracts. For EOSIO smart contract vulnerability detection, it can detect the same vulnerabilities as EOSFuzzer: fake EOS attacks, fake transfer notifications, and block dependency.\par

\textbf{EOSafe} is a static analysis framework for automatically detecting vulnerabilities in EOSIO smart contracts at the bytecode level. It takes the Wasm bytecode of EOSIO smart contracts as input and, like EVulHunter, is based on Octopus. EOSafe can detect four types of vulnerabilities: fake EOS attacks, fake transfer notifications, rollback attacks, and permission control vulnerabilities.\par

All four tools can produce false positives or false negatives. False positives occur when a contract is reported to have a vulnerability that it does not actually have, while false negatives occur when a contract with an actual vulnerability is not detected correctly.\par

\section{Behavioral Security}
The quality of a blockchain system should be evaluated comprehensively from multiple aspects, including architecture, performance, and security. In previous sections, we have analyzed the architecture, decentralization, performance, and smart contract security of the EOS system. However, attacks on the EOS blockchain may exploit not only vulnerabilities in smart contracts but also vulnerabilities in the EOS system architecture itself. EOSIO's unique system design, while enhancing transaction throughput and scalability, also presents some potential security risks. This section will introduce four possible attacks targeting EOSIO and provide an analysis of real-world attack incidents in DApps.\par

\subsection{Possible Attacks on EOSIO}
\subsubsection{Block Delay Attack}
Block production delay attack: A block production delay attack targets block producers, where malicious nodes send a large number of unprocessable fake transactions to disrupt the normal operation of block producers. These fake transactions occupy the processing opportunities for legitimate transactions, causing the legitimate transactions to be delayed and remain unprocessed, resulting in significant economic losses.\par

In the EOS system, block producers have four states. Here, we will introduce two of them: the "success" state and the "exhausted" state. Since the resources such as CPU, NET, and RAM available to block producers are limited, there are restrictions on the resources allocated for transaction processing within a block. If the block producer can complete processing within 0.5 seconds, it enters the waiting time and sends the block to other producers. This is the "success" state. If the resources required for processing a block exceed the specified threshold, the block producer enters the "exhausted" state. In this state, the producer will try to produce as many blocks as possible, and even if it completes processing within 0.5 seconds, it will not enter the waiting state to maximize resource usage. A block production delay occurs when the block producer transitions from the exhausted state, after producing a large number of blocks, back to the success state.\par

Two features of EOSIO can be used to implement block delay attacks: delayed transactions and smart contract updates. Normally, block producers prioritize processing requested transactions and delay the execution of deferred transactions. If an attacker deploys a malicious contract that recursively calls its deferred transactions, the number of transactions will grow exponentially. When the block producer receives such a massive number of transactions, its resources will be exhausted. After producing a large number of blocks, the attacker updates the malicious contract to a completely different one. This causes the transactions from the original malicious contract to become invalid. When the block producer processes these invalid transactions, more blocks are generated. This process causes a discrepancy between the block timestamps and real time, leading to a block delay attack.\par

\subsubsection{CPU Exhaustion Attack}
CPU exhaustion attack: The execution of smart contracts consumes a certain amount of CPU power, and the consumption is determined by the execution time of the smart contract. A CPU exhaustion attack refers to an attacker depleting the target's CPU power to the extent that the smart contract can no longer be executed. The goal of the attacker is to exhaust the contract owner's CPU resources, preventing the contract from being called by block producers, thus rendering the smart contract unavailable.\par
Smart contracts often use the `send\_deferred()` function to generate deferred transactions. The execution of such smart contracts requires CPU resources from the transaction initiator or the contract owner. Contract owners often provide their CPU resources to ensure the widespread use of their smart contracts. Attackers take advantage of this by continuously consuming the contract owner's CPU resources, causing the owner's CPU to be depleted. As a result, the contract will be unable to execute, effectively making the smart contract unavailable.\par

\subsubsection{RAM Exhaustion Attack} 
The RAM exhaustion attack is similar to the CPU exhaustion attack, with the key difference being that RAM is used to store data. In this case, an attacker will send a large amount of fake data to the contract owner, aiming to deplete the owner's RAM. The defense against this type of attack is to prevent the same user from storing unlimited data.

\subsubsection{RAMsomware Attack}
In a ransomware attack, the attacker can seize all of the victim's resources, including CPU, RAM, and EOS tokens. The attacker takes advantage of the contract owner's sensitive permissions. Initially, the attacker uses a normal contract to request the `eosio.code` permission from the user to proceed with further actions. The victim, after verifying the contract's security through a third-party inspection, grants the contract the necessary permissions. Once the smart contract passes the inspection, the attacker can modify it into a malicious contract and exploit the granted sensitive permissions to launch the attack. The victim will not realize that the contract has been altered.

\subsection{Reality Attack Case Analysis}

\subsubsection{EOS Blockchain Real-World Attack Events}
This section reviews notable real-world attack events on the EOS mainnet since its launch, detailing the background and aftermath of these incidents.\par
\textbf{EOS Fomo3D Werewolf Game}: This game was similar to the Fomo3D game on Ethereum. On July 25, 2018, the SlowMist Security Team issued a warning about an overflow attack on the EOS Fomo3D game contract, which caused the fund pool to go negative. In response, the Werewolf team took emergency measures and launched a new contract, but another attack followed, where an attacker (eosfomoplay1) stole 60,686 EOS. \par

\textbf{EOSBet}: EOSBet is a dice-based gambling game on the EOS blockchain. It suffered three attacks. On August 26, 2018, the project team reported that its account's RAM was maliciously consumed by a contract. The game temporarily shut down to prevent further attacks. On August 27, the vulnerability was fixed, and the game resumed. However, on September 14, hackers exploited the lack of verification of whether the received EOS was issued by `eosio.token`, resulting in a loss of 44,427.4302 EOS and 1,170.0321 BET tokens. Despite transferring assets to a cold wallet and upgrading security, the project suffered another attack on October 15, when the attacker (ilovedice123) used the "fake transfer notification" vulnerability, causing a loss of 142,845 EOS. \par

\textbf{Other Gambling Platforms}: Several EOS-based gambling platforms, such as Luckyos, EOS WIN, and DEOSGames, also suffered from random number attacks. For example, on August 27, 2018, the random number generation for Luckyos' rock-paper-scissors game was cracked, leading to its shutdown. Similarly, on September 2, 2018, the random number for EOS WIN was compromised, resulting in a 2,000 EOS loss. 

\textbf{EOSBank}: EOSBank, a platform supporting EOS deposits and leasing services, was attacked on October 5, 2018. The attacker modified the owner permissions of the EOSBank contract account `eosiocpubank`, resulting in the transfer of 18,000 EOS to the account `fuzl4ta23d1a`. \par

\subsubsection{DApp Transfer Process and Attack Analysis}
EOS offers two methods for implementing DApp transfers:
\textbf{User Authorization of `eosio.code` Permission to DApp}: In this approach, the DApp can execute the `transfer` function of `eosio.token` to transfer EOS on behalf of the user.\par
\textbf{Direct Call to `eosio.token`'s Transfer Function}: In this method, the user directly calls `eosio.token`'s `transfer` function to transfer EOS to the DApp account. This method is preferred due to security risks in the first method, where granting `eosio.code` permission gives DApp complete control over the user’s account.\par

The transfer process involves five stages. The analysis of potential attack methods is as follows: \par

Stage 1 (System Contract Logic): In this stage, the transfer process follows the logic of the `eosio.token` system contract. If the DApp is correctly deployed, there are no exploitable vulnerabilities at this stage.
Stage 2 (Apply Phase): This phase could be vulnerable to a fake EOS attack. In the previous section, the fake EOS attack was explained. Essentially, a malicious contract simulates the issuance of fake EOS tokens. The attack occurs when the malicious contract sends fake EOS tokens to the DApp, bypassing the detection logic since the only check in the transfer function is the token name ("EOS"), which is easily bypassed by using fake EOS tokens. Stage 3 (Transfer Phase): In this phase, a fake transfer notification attack could occur. In this attack, an attacker sends real EOS tokens to their own smart contract account. The `eosio.token` contract calls `require\_receipt`, which triggers the malicious smart contract’s `transfer` function. This results in the malicious contract calling the DApp’s `transfer` function, making it appear as though a legitimate transfer took place. Stages 4 \& 5 (Receipt and Reveal Phases): These phases are vulnerable to random number attacks. In these stages, DApp's internal logic is executed, and attackers aim to predict or manipulate the random number generated. Many DApps use pseudo-random numbers, which can be vulnerable to attacks. Attackers can predict the random number if they control the `ref\_block\_num` (which is user-defined). By choosing a specific `ref\_block\_num`, attackers can compute the random number in advance, allowing them to manipulate the results. To mitigate this, DApps can use future data as random numbers by implementing deferred actions, as shown in the provided process flow.\par
This analysis underscores the importance of proper validation and security measures in the design of DApp transfer processes to prevent such vulnerabilities.

\section{Conclusion}
This paper provides a comprehensive analysis of the EOS blockchain system from several aspects, including system architecture, decentralization, performance, smart contracts, and behavioral security. In terms of system architecture, the paper analyzes the EOS system architecture from six core components, detailing the characteristics and workings of each component. In the area of decentralization, the analysis focuses on accounts, the supernode election process, and block producers. The study reveals issues such as a low number of active accounts, limited participation in the election process, and little variation in the set of block producers within the EOS blockchain. In performance, the paper examines EOS block production, transaction and operation statistics, and resource utilization. It concludes that the performance of the EOS system, while influenced by multiple factors in real-world use, does not exceed Ethereum's by a significant margin, though its potential remains promising. Regarding smart contracts, the paper introduces five known contract vulnerabilities and compares four vulnerability detection platforms. Finally, in terms of behavioral security, the paper summarizes four types of attacks that EOS may face due to its architectural design.
\vspace{10pt}
\bibliographystyle{IEEEtran}
\bibliography{main} 
\end{document}